\documentstyle[preprint,aps,prl,epsf]{revtex}
\begin{document} 
\setcounter{footnote}{0}
\bibliographystyle{plain}
\draft
\title{Scalar-relativistic effects in solids in
the framework of a Douglas--Kroll transformed 
Dirac--Coulomb Hamiltonian}
\author{{\bf{Norbert J.M.\ Geipel, Bernd A.\ He{\ss}}}}
\address{Institut f\"ur Physikalische und Theoretische Chemie, \\
Wegelerstr.\ 12, Universit\"at Bonn, D-53115 Bonn, Germany}
\date{Received 11 April 1997}
\maketitle
\newcommand{\kjmol}{\textstyle{\rm{kJ\,mol^{-1}}}}
\newcommand{\cmmo}{\textstyle{\rm{cm^{3}\,mol^{-1}}}}
\begin{abstract}
The Douglas--Kroll transformed Dirac--Coulomb
Hamiltonian is used to describe scalar-relativistic
effects in solids.\ A Hartree--Fock
approximation with periodic boundary conditions
makes it feasible to use methods originally developed
for atoms and molecules to solve the corresponding
equations for crystalline systems.\ The implementation
is realized within the CRYSTAL program.\ 
Scalar-relativistic effects in silver compounds of FCC structure
such as AgF, AgCl, AgBr and Ag are investigated for the molar
formation energy, the lattice parameter, the isothermal 
bulk modulus and the pressure derivative.
\end{abstract}
\setcounter{section}{0}
\section*{1.\ Introduction}
In the last decade, methods for treating
relativistic effects in heavy atoms and molecules have been
developed in order to extend the applicability of
ab initio quantum chemical methods to heavy atoms 
and molecules.\
Recently, different approaches and their realizations were
compared as concerning the accuracy and efficiency of the method
\cite{collins}.\ A detailed discussion 
of the accuracy of different relativistic
approximations by the investigation of hydrogenic
energy levels can be found in Ref.\ \cite{schwarz}.\
The Douglas--Kroll (DK) transformed 
Dirac--Coulomb Hamiltonian \cite{douglas,hess1} 
is considered to be 
an efficient and reliable 
approximation for calculations on relativistic systems.\ 
The theoretical 
framework of this approximation is based on quantum
electrodynamics, and as one main consequence this
relativistic Hamiltonian is variationally stable and therefore
has normalizable bound-state solutions.\
The application of 
external-field projection operators which project
onto the positive-energy states of the Dirac
single-particle Hamiltonian
leads to a two-component formalism which is separable
in a spin-free and a spin-dependent part.\ There has
been an extensive discussion of relativistic effects
in the framework of this approach for
atoms and molecules such as
silver and gold \cite{samzow,kaldor},
gold hydride \cite{kaldor,jansen} and gold fluoride
\cite{schwerdtfeger}
using the Hartree--Fock
(HF) approximation as well as
various methods including electron correlation effects
and spin--orbit effects \cite{marian2}.\

There is also a plenitude of methods for treating
relativistic effects in solids.\
The mainstream approach
in this field being density functional theory (DFT) in the
local density approximation (LDA).\ All these methods are, in
principle, based on a relativistic extension of the 
Hohenberg--Kohn theorem \cite{ramana}.\ Most of these
approaches rely on the four-component treatment of a 
spherical symmetric potential (see Refs.\ [11,12] and references
therein), which can be relatively
easily solved using appropriate methods from atomic theory,
extended to effective one-particle equations given by
MacDonald and Vosko \cite{vosko}.\ One possibility to elude
the shape approximation for the crystal potential are
relativistic
effective-core potentials (ECP) which are adjusted to 
(quasi-) relativistic atomic calculations (see Ref.\ [14] and
references therein) and have to be transferable to the 
corresponding crystalline situation
\cite{bachelet,christensen,elsaesser}.\

The extension of
HF methods in solid-state theory
to the relativistic case \cite{ladik} is less obvious, however
in practice, the treatment of potentials of non-spherical
symmetry (and the non-local exchange potential)
requires an effort which is much larger than in the non-relativistic
case.\
Consequently, it seems to be a reasonable idea to use
the DK transformation for the one-electron
operators also in HF calculations with periodic
boundary conditions.\ In this Letter we describe a corresponding
implementation of the method in 
the CRYSTAL88 program \cite {dovesi2,pisani}
and report first applications of the code.\

\section*{2.\ Method}
The spin-free part of the DK
Hamiltonian is used
to treat scalar-relativistic
effects in solids.\ Using atomic units (with the electron
mass $m=1$), the relativistic
operator has the following form:
\begin{eqnarray}
H^{\rm{sf}_{1}}_{+}&=&\sum_{i}\:[\: E_{i}-m\,c^{2}
+V^{\rm{sf}}_{\rm{eff}}(i)\:]
\:+\sum_{i<j}\:\frac{1}{r_{ij}},
\nonumber \\[1.0ex]
E_{i}&=&(p^{2}_{i}\,c^{2}+m^{2}\,c^{4})^{\:\scriptstyle{\frac{1}{2}}},
\nonumber \\[1.0ex]
V^{\rm{sf}}_{\rm{eff}}(i)&=&-A_{i}\, [\, V_{\rm{ext}}(i)
+{\bf{R}}_{i}\, V_{\rm{ext}}(i)\, {\bf{R}}_{i}\, ]\, A_{i}
- {\textstyle{\frac{1}{2}}}\:\{\{E_{i},W_{1}^{\rm{sf}}(i)\} ,
W_{1}^{\rm{sf}}(i)\}, \label{relop} \\[1.0ex]
\mbox{with  } A_{i} &=& \sqrt{\frac{m\,c^{2}\,+\,E_{i}}{2 E_{i}}}
\mbox{,  } {\bf{R}}_{i} = \frac{c\, {\bf{p}}_{i}}
{m c^{2}\,+\,E_{i}}.
\nonumber
\end{eqnarray}
\noindent
The brackets $\{,\}$ denote the anticommutator $\{a,b\} = ab + ba$.\
The operator $W^{\rm{sf}}_{1}$ is an integral operator with kernel
\begin{eqnarray}      
{\cal{W}}^{\rm{sf}}_{1}({\bf{p}}_{i},{\bf{p}}_{i}^{'})=
A_{i} ({\bf{R}}_{i}-{\bf{R}}_{i}^{'}) A_{i}^{'}
\frac{V_{\rm{ext}}({\bf{p}}_{i},{\bf{p}}_{i}^{'})}
{E_{i}+E_{i}^{'}}. \nonumber     
\end{eqnarray}
\noindent Within this approach the use of the non-relativistic
Coulomb
repulsion instead of the transformed two-electron term
is a good approximation
if scalar-relativistic effects are considered \cite{samzow}.\

An all-electron ab initio HF approximation for crystalline
systems has been realized in the CRYSTAL88 program 
\cite{dovesi2,pisani}.\ To construct
a Fock operator in
an atomic orbital (AO) basis (direct space
representation) within this approach
including the above relativistic operators, three
main steps are necessary:

\noindent 1.\ In order to represent the operator ${\bf{R}}_{i}\,
V_{\rm{ext}}(i)\,{\bf{R}}_{i}$ in Eq.\ (\ref{relop}) in an AO basis,
we have to calculate matrix elements of the operator
${\bf{p}}_{i}\,{V_{\rm{ext}}}(i)\,{\bf{p}}_{i}$.\

\noindent 2.\ The Bloch function (BF) representation
of the kinetic energy,
the external potential and the additional matrix elements
mentioned above is used
to get a representation of the relativistic operators in
the BF basis.\

\noindent 3.\ To get a representation of the relativistic operators
in direct space, those quantities are Fourier transformed back to
the AO basis.\

We make use of the fact that scalar-relativistic
effects are short-range \cite{christensen}.\
Therefore, the long-range tail of the 
electrostatic potential calculated in the CRYSTAL88 program 
by means of Ewald
sums \cite{saunders}
is not 
relativistically corrected.\
Rather, the relativistic contribution of the
external potential described by Eq.\ (\ref{relop}) to the Fock operator
is restricted to a finite number of direct lattice vectors ${\bf{h}}$
within the spatial region that is treated
by the
direct lattice series of the Ewald potential 
function \cite{saunders} -- hereafter, dubbed 
"direct space potential zone".\ 

Each AO is expanded in real spherical Gaussian-type functions (SGTF)
\cite{pisani,saunders2}.\
We denote by $[\omega \,{\bf{g}}] =
{\chi}_{\Omega}({\bf{r}} -
{\bf{g}} - s_{\omega})$ the $\omega$th AO of the
primitive unit cell identified by the vector
${\bf{g}}$ located on an atom whose image in the 
primitive unit cell is assigned by the fractional vector
$s_{\omega}$;
${\bf{g}}, {\bf{h}}, ...$ label direct lattice
vectors
and $\Omega = (n,l,m)$ denotes
the quantum numbers characterizing the AO.\
For simplicity
we write 1,2 for $\omega_{1}$, $\omega_{2}$
in the following.\
To evaluate the integrals, we
calculate additional matrix elements of
$V^{C}_{{\rm{ext}}}({\bf{r}}) =  
\sum^{C}_{{\bf{h}}}\,
\sum_{a}\,Z_{a}\,|{\bf{r}}\,-\,{\bf{h}}\,-\,{\bf{s}}_{a}|^{-1}$ 
in a basis of SGTF of the form
\begin{eqnarray}
R(12{\bf{g}}) = <\nabla 1 {\bf{O}} | V^{C}_{ext} ({\bf{r}})
| \nabla 2 {\bf{g}} >. \nonumber
\end{eqnarray}
These integrals result from the operator
${\bf{R}}$, which is equal to the linear momentum
${\bf{p}} = -i \nabla$ \\ times a $p^{2}$-dependent 
kinematical factor.\
In practice, we                     
calculate these derivatives with Cartesian Gaussian-type functions
(CGTF)
and transform
the corresponding integrals to
SGTF afterwards \cite{saunders2}.

In order to adapt the basis set
to the translational invariance of the crystal, the AOs 
are used to build BFs \cite{pisani,dovesi}.
It is convenient to use this basis
to get a matrix representation of non-local operators
such as $E$, $A$ and $(E\,+\,mc^{2})^{-1}$ because
by definition this is the basis in which the matrix
of one-electron operators is symmetry blocked.\ When the BF 
representation of the
kinetic energy operator ${\widehat{T}}$
\begin{eqnarray}
{\widehat{T}}(12{\bf{k}}) = S(12{\bf{k}})^{-1/2}\:
T(12{\bf{k}})\:
S(12{\bf{k}})
^{-1/2} \nonumber
\end{eqnarray}
is expressed in its eigenbasis
\begin{eqnarray}
&& {\widehat{T}}(12{\bf{k}}) = U(12{\bf{k}})\: 
\lambda(12{\bf{k}})\:
U^{\dagger} (12{\bf{k}})
\nonumber
\end{eqnarray}
\noindent --- the diagonal matrix
$\lambda (12{\bf{k}})$ contains the positive real 
eigenvalues of ${\widehat{T}}(12{\bf{k}})$
and the unitary matrix $U(12{\bf{k}})$ the corresponding eigenvectors
---
the BF representation of an analytic
function $f({\widehat{T}}(12{\bf{k}}))$ 
corresponding to operators
like ${\widehat{E}}(12{\bf{k}})$,
${\widehat{A}}(12{\bf{k}})$ and
$({\widehat{E\,+\,mc^{2}}})^{-1}(12{\bf{k}})$
is given by
\begin{eqnarray}
&& f({\widehat{T}}(12{\bf{k}})) = 
U(12{\bf{k}}) \: f(\lambda(12{\bf{k}})) \:
U^{\dagger} (12{\bf{k}}). \nonumber
\end{eqnarray}
\noindent Here we assume that this relationship holds
approximately also in the finite matrix representation
given by the AO basis set, from which the BFs
are constructed.\
The contribution of those operators 
to the relativistic external potential $V^{C,\rm{sf}}_{\rm{eff}}
(12{\bf{k}})$ given by Eq.\ (\ref{relop}) --- restricted
to the direct space potential zone ---
is then realized by simple matrix
multiplication for every ${\bf{k}}$ point within the 
irreducible wedge of the first Brillouin
zone (IBZ).\

The relativistic external potential
$V^{C,\rm{sf}}_{\rm{eff}}(12{\bf{k}})$ transforms
like a totally symmetric operator.\ To transform this operator
back to the AO basis, the Fourier transformation of
$V^{C,\rm{sf}}_{\rm{eff}}(12{\bf{k}})$
restricted 
to the IBZ (see Refs.\ [20,23] and references therein)
generates 
a matrix
\begin{eqnarray}
\Pi^{C,\rm{sf}}_{\rm{eff}}(12{\bf{g}}) = \int_{IBZ}\:
V^{C,\rm{sf}}_{\rm{eff}}
(12{\bf{k}})\,\exp(i\,{\bf{k}}\cdot{\bf{g}})\,d{\bf{k}} \nonumber
\end{eqnarray}
\noindent from which the full potential matrix 
$V^{C,\rm{sf}}_{\rm{eff}}(12{\bf{g}})$ is generated by
applying the appropriate rotation matrices \cite{pisani,dovesi}.\
The relativistic operator $E(12{\bf{k}})$ is treated likewise.\

The Fock operator expressed
in an AO basis can
formally be written as sum of a
one-electron and a two-electron part
\begin{eqnarray}
F(12{\bf{g}}) = H(12{\bf{g}})\,+\,B(12{\bf{g}}). \nonumber
\end{eqnarray}
The contribution of Eq.\ (\ref{relop}) to the Fock operator
is related to $H(12{\bf{g}})$  and
does not include any relativistic correction 
of the long-range behaviour of the electrostatic
potential $Z(12{\bf{g}})$ because it is restricted to
the direct space potential zone.\
We thus include relativistic effects due to the external
potential by adding the difference of the relativistic
potential and the non-relativistic potential
\noindent calculated in this zone to the total non-relativistic
potential as calculated in the CRYSTAL88 program.\
\begin{eqnarray}
H(12{\bf{g}}) = E(12{\bf{g}})\,+\,V^{C,{\rm{sf}}}_{\rm{eff}}
(12{\bf{g}})\,-\,V^{C}_{\rm{ext}}(12{\bf{g}})
\,+\,Z(12{\bf{g}}). \nonumber
\end{eqnarray}
Thus,
the delicate treatment of the Coulomb series 
\cite{saunders}
remains the same.\ As usual the total
HF energy per primitive unit cell is given by
\begin{eqnarray}
\epsilon = {\displaystyle{\frac{1}{2}}}\sum_{1,2,{\bf{g}}}\,
P(12{\bf{g}})\,(2 H(12{\bf{g}})\,+\,B(12{\bf{g}})), \nonumber
\end{eqnarray}
\noindent where $P(12{\bf{g}})$ denotes the density matrix
of a crystal \cite{pisani2}.

\section*{3.\ Technical details}
Scalar-relativistic effects in several simple silver compounds 
of FCC structure such as AgF, AgCl, AgBr and Ag are investigated using 
this method with the reoptimized uncontracted basis sets for F, Cl and Br
of Dunning \cite{dunning} and for Ag of
Martin \cite{martin}.\
In crystalline compounds low-exponent functions are likely
to introduce linear dependencies.\ 
Therefore, diffuse exponents ($\leq 0.04$) of the basis set for Ag
are discarded.\ 
Concerning the question whether HF solutions for conductors are
appropriate in spite of the vanishing density of states at the
Fermi level and the related problem of finding appropriate HF solutions of 
the free electron gas, we refer to Ref.\ [24] and references
therein.\
Basis sets
and details of the atomic and the CRYSTAL88 HF calculations 
using $H^{{\rm{sf}}_{1}}_{+}$ in the relativistic case and
resulting total HF energies are given in Tables \ref{tab0} and \ref{tab1}.\ 

All calculations are realized
close to program limitations and whereever possible
parameters for high-precision calculations \cite{pisani} are used.\
To realize an accurate representation of the relativistic operators,
it is convenient to use 
relatively large uncontracted basis sets.\
In order to obtain a reliable transformation
from indirect space (${\bf{k}}$ space) to direct space,
the number
of ${\bf{k}}$ points used to realize the integration
over the IBZ should also be large.\
Testing the method described above 
for $T(12{\bf{k}})$ and $V^{C}_{\rm{ext}}(12{\bf{k}})$
by comparing its IBZ integration with the
given integrals of $T(12{\bf{g}})$ and $V^{C}_{\rm{ext}}
(12{\bf{g}})$, it turned out that
29 ${\bf{k}}$ points of the Monkhorst net \cite{pisani} within the IBZ 
are sufficient to obtain
an absolute error for the integrals which is less than 
$10^{-8}$.\
Moreover, the change in the total HF energy per primitive unit cell
for metallic silver is less than
$0.9\times10^{-4}$ au
if we use 72 ${\bf{k}}$ points instead of 29 ${\bf{k}}$ points
within the IBZ
for the relativistic version of the program.

The geometry optimization is realized with the restart option
GEOM of the CRYSTAL88 program \cite{pisani}.\
Two different equations of state (EOS) deduced by Murnaghan 
and Birch, respectively, \cite{murn} are used to fit 
the total energy as a function of the lattice parameter.\
Though nearly all the results
are only weakly dependent on the applied EOS 
--- except the pressure derivatives
of Ag and AgBr --- the accuracy of the fit
and especially related results such as bulk moduli and pressure
derivatives have to be investigated carefully (for details see
Table \ref{tab2}).\ 

A corresponding ECP calculation  
of AgCl
\cite{apra1} is repeated (see Table \ref{tab3}) to
ensure compatibility of all used parameters.\ 
Since small-core pseudopotentials for Ag are used, it is
assumed that frozen-core effects are small and relativistic
all-electron HF
calculations are comparable with corresponding ECP
calculations.\

\section*{4.\ Results}
Although relativistic effects are widely discussed in literature
(see Ref.\ \cite{pyykkoe} and references therein),
ab initio calculations of silver halides with
FCC structure including scalar-relativistic
corrections are rare
\cite{wood} and in most cases pseudopotential approaches are used
to consider those effects (see Ref.\ \cite{apra1} and
references therein).\ By way of contrast,
due to the simple structure of diamagnetic
metals relativistic calculations 
of metallic silver are numerous (see Refs.\ [16,17] 
and references therein).\ 
Nevertheless, scalar-relativistic effects without any substantial
approximation of the crystal potential have not been
investigated before.\

The results of the investigated FCC crystals are collected in 
Table \ref{tab2}.\  Properties such as the lattice parameters and the
related molar volumes show small effects --- they decrease up 
to 2.0\% respectively up to 5.8\% (for Ag)
as compared to the non-relativistic values.\
The corresponding decrease of the nearest-neighbor distances
lies between 0.8 pm for AgF and 6.4 pm for Ag ---
relativistic bond-length contractions of similar molecules
are larger \cite{pyykkoe,ziegler}.\ Looking at the molar HF binding
energies, a remarkable bond destabilization of the halides is observed
--- the molar HF binding energies are higher up to 8.9\% (for
AgF) as compared to the non-relativistic values.\  This trend is in
accord with the finding in molecular compounds that noble-metal
halides with a halogen more electronegative than the metal itself
experiences a relativistic destabilization of the bond, along with a
bond length contraction \cite{schwerdt}.\
For Ag
we find a large bond stabilization of about 34.5\%.\
Bulk moduli and pressure derivatives change up to
7.5\% (for Ag) respectively up to 8.2\% (for AgCl)
when the EOS of Murnaghan \cite{murn} is used.\ Corresponding
fits of total HF energies versus the lattice parameters
of AgBr and Ag are shown in Fig.\ 1 and Fig.\ 2.,
respectively.\

Although scalar-relativistic HF calculations are not directly
comparable with calculations using scalar-relativistic
ionic pseudopotentials in the framework of the LDA \cite{elsaesser}
due to the missing treatment of
correlation effects, our results
suggest that relativistic effects in 4d metals are
overestimated when ionic pseudopotentials are used
(for a comparison of absolute values see annotation in Table \ref{tab2}).\
This could explain why
non-relativistic values of 4d metals 
are slightly closer to experiment than scalar-relativistic
ones (see Ref.\ \cite{elsaesser} and references therein).\

In Table \ref{tab3} we compare crystal properties of AgCl at various
levels of theory.\ All calculations are founded upon a periodic
HF approach realized in the CRYSTAL program 
\cite{dovesi2,crystal92}.\ Results for different a posteriori
density functional corrections and experimental values 
are included \cite{apra1}.\
We find remarkable
agreement between relativistic HF calculations using 
$H^{{\rm{sf}}_{1}}_{+}$ and relativistic ECP calculations
using pseudopotentials of Hay and Wadt (see Ref.\ [29])
concerning the bulk modulus and the
lattice parameter.\ Slight differences are observed for
the molar formation energy and the pressure derivative.\
Corresponding fits of total HF energies per primitive
unit cell
versus the lattice parameter of AgCl are shown in Fig.\ 3.

\section*{5.\ Conclusions}
We have calculated scalar-relativistic effects of
crystalline properties such as
lattice parameters, molar formation energies,
bulk moduli and pressure derivatives of various silver 
compounds with FCC structure by means of an implementation
based on the Douglas--Kroll transformation 
within a periodic Hartree--Fock approach realized in the
CRYSTAL program.\ 

Good agreement between a relativistic
effective-core potential calculation and
the one-component Douglas--Kroll
approach is obtained for most of the properties 
investigated for 
AgCl.\ Though scalar-relativistic effects of
silver compounds are expected to be small,
remarkable changes of the Hartree--Fock
binding energy and of the bulk modulus are
observed.

\section*{Acknowledgements}
The support of the "Deutsche Forschungsgemeinschaft"
in the framework of the
"Sonderforschungsbereich 334: Wechselwirkungen in Molek\"ulen"
and of the European Science Foundation within the REHE program are
gratefully acknowledged.\

\pagebreak

\small{
\pagestyle{empty}
\begin{table}

\vspace*{6.0ex}
\caption{\label{tab0} Basis sets used for relativistic$^{\rm{a}}$
and  non-relativistic atomic Hartree--Fock (HF) calculations for
the ground state of Ag, F, Cl and Br
and their total
HF energies ($E^{\rm{tot}}$)}
\begin{tabular}{clll|lcc}
Atom&\multicolumn{3}{c|}{Outermost exponents$^{\rm{b}}$}&
Method&State& $E^{\rm{tot}}$ \\
    &\multicolumn{1}{c}{s}
    &\multicolumn{1}{c}{p}
    &\multicolumn{1}{c|}{d\hspace*{2.0ex}}
& & & ($E_{\rm{h}}$) \\
\tableline
Ag&\multicolumn{6}{l}{Original basis: (15s10p8d) [26]
\hspace{1.0ex}
$\longrightarrow$\hspace{1.0ex}
Final basis: (13s9p8d)} \\
  &\multicolumn{5}{l}{Exponents deleted:\hspace{1.0ex}
0.04 (s) \hspace{0.5ex} 0.015 (s) \hspace{0.5ex} 0.025 (p)}\\
\tableline
& & & &HF limit [35]&$^{2}{\rm{S}}$&
$-5197.6985$ \\
&0.74 (0.75)&0.81&0.40 (0.39)\hspace*{2.0ex}&nonrel.\ HF&
$^{2}{\rm{S}}$&$-5197.0041$ \\
&0.06\hspace*{7.0ex}&0.29&0.11 (0.12)\hspace*{2.0ex}
&rel.\ HF$^{\rm{a}}$&
$^{2}{\rm{S}}$&$-5310.3549$ \\
\tableline
F&
\multicolumn{5}{l}{Original basis: (9s5p) [25]} \\
\tableline
& & & &HF limit [35]&$^{2}{\rm{P}}$&$-99.4094$ \\
&1.21&0.94& &nonrel.\ HF&
$^{2}{\rm{P}}$&$-99.3955$ \\
&0.36\hspace*{1.0ex}&0.27&&rel.\ HF$^{\rm{a}}$&
$^{2}{\rm{P}}$&$-99.4821$ \\
\tableline
Cl&
\multicolumn{5}{l}{Original basis: (11s7p) [25]} \\
\tableline
& & & &HF limit [35]&$^{2}{\rm{P}}$&$-459.4821$ \\
&0.53&0.64\hspace*{1.0ex}& &nonrel.\ HF&
$^{2}{\rm{P}}$&$-459.4382$ \\
&0.19&0.18& &rel.\ HF$^{\rm{a}}$&
$^{2}{\rm{P}}$&$-460.8422$ \\
\tableline
Br&
\multicolumn{5}{l}{Original basis: (14s11p5d) [25]} \\
\tableline
& & & &HF limit [35]&$^{2}{\rm{P}}$
&$-2572.4413$ \\
&0.43\hspace*{1.0ex}&0.47\hspace*{1.0ex}&3.851\hspace*{2.0ex}&
nonrel.\ HF&
$^{2}{\rm{P}}$&$-2572.3577$ \\
&0.16&0.14&1.32\hspace*{2.0ex}&rel.\ HF$^{\rm{a}}$&
$^{2}{\rm{P}}$&$-2604.1851$ \\
\end{tabular}

\vspace{2.0ex}

\noindent $^{\rm{a}}$ Calculations using $H^{{\rm{sf}}_{1}}_{+}$.

\noindent $^{\rm{b}}$ Only reoptimized exponents (au) are
tabulated.\ Exponents which differ for the relativistic 
calculation are given in parentheses.

\pagebreak

\caption{\label{tab1} Basis sets used for relativistic$^{\rm{a}}$
and non-relativistic CRYSTAL88 Hartree--Fock (HF) calculations
of the FCC crystals Ag, AgF, AgCl and AgBr
and their total
HF energies ($E^{\rm{tot}}$) with reference to the
primitive unit cell}
\begin{tabular}{clll|clcc}
Basis set&\multicolumn{3}{c|}{Outermost exponents$^{\rm{b}}$}
&FCC crystal&\multicolumn{1}{l}
{Method}&Accuracy&$E^{\rm{tot}}$ \\
for &\multicolumn{1}{c}{s\hspace*{4.0ex}}
&\multicolumn{1}{c}{p}
&\multicolumn{1}{c|}{\hspace{2.0ex}d}
& & &parameter$^{\rm{c}}$& ($E_{\rm{h}}$) \\
\tableline
Ag&\multicolumn{7}{l}{Original basis: (15s10p8d)
\hspace{1.0ex}
$\longrightarrow$\hspace{1.0ex}
Final basis: (13s9p8d) [see Table I]} \\
\tableline
Ag&0.77&0.805&\hspace*{3.0ex}0.385\hspace*{2.0ex}&Ag&nonrel.\ HF&
6 8 7 7 13& $-5197.0191$ \\
&0.10&0.22\hspace*{1.0ex}&\hspace*{3.0ex}0.13\hspace*{1.0ex}&
Ag&rel.\ HF$^{\rm{a}}$&
6 8 7 7 13&$-5310.3751$ \\
\tableline
Ag&\multicolumn{7}{l}{Original basis: (15s10p8d)
\hspace{1.0ex}
$\longrightarrow$\hspace{1.0ex}
Final basis: (13s9p8d) [see Table I]} \\
F&\multicolumn{6}{l}{Original basis: (9s5p) [25]} \\
\tableline
Ag&0.77&0.78&\hspace*{3.0ex}0.43\hspace*{1.0ex}&&&
  & \\
&0.08&0.24&\hspace*{3.0ex}0.14\hspace*{1.0ex}&
AgF&nonrel.\ HF&6 8 6 6 13
&$-5296.5548$ \\
F&1.25&0.95& 
&AgF&rel.\ HF$^{\rm{a}}$&6 8 6 6 13
&$-5409.9784$ \\
&0.38&0.28& & & &
 & \\
\tableline
Ag&\multicolumn{7}{l}{Original basis: (15s10p8d)
\hspace{1.0ex}
$\longrightarrow$\hspace{1.0ex}
Final basis: (13s9p8d) [see Table I]} \\
Cl&\multicolumn{7}{l}{Original basis: (11s7p) [25]
\hspace{1.0ex}
$\longrightarrow$\hspace{1.0ex}
Final basis: (11s7p1d)} \\
\tableline
Ag&0.79&0.78&\hspace*{3.0ex}0.40\hspace*{1.0ex}&&&
  & \\
&0.11&0.24&\hspace*{3.0ex}0.12\hspace*{1.0ex}&AgCl&
nonrel.\ HF&6 8 6 6 13
&$-5656.6014$ \\
Cl&0.54&0.70&
&AgCl&rel.\ HF$^{\rm{a}}$&6 8 6 6 13
&$-5771.3441$ \\
&0.20&0.20&\hspace*{3.0ex}0.14\hspace*{1.0ex}&
&&& \\
\tableline
Ag&\multicolumn{7}{l}{Original basis: (15s10p8d)
\hspace{1.0ex}
$\longrightarrow$\hspace{1.0ex}
Final basis: (13s9p8d) [see Table I]} \\
Br&\multicolumn{6}{l}{Original basis: (14s11p5d) [25]} \\
\tableline
Ag&0.76&0.76&\hspace*{3.0ex}0.40\hspace*{1.0ex}&&&
  & \\
&0.12\hspace*{4.0ex}&0.11&\hspace*{3.0ex}0.12\hspace*{1.0ex}&AgBr&
nonrel.\ HF&6 8 7 7 13
&$-7769.5003$ \\
Br&0.46\hspace*{4.0ex}&0.57&\hspace*{3.0ex}3.851\hspace*{0.0ex}
&AgBr&rel.\ HF$^{\rm{a}}$&6 8 7 7 13
&$-7914.6672$ \\
&0.16\hspace*{4.0ex}&0.19&\hspace*{3.0ex}1.30\hspace*{1.0ex}& & &
 & \\
\end{tabular}
 
\pagebreak

\noindent $^{\rm{a}}$ Calculations using $H^{{\rm{sf}}_{1}}_{+}$.

\noindent $^{\rm{b}}$ Only reoptimized exponents (au)
are tabulated.

\noindent $^{\rm{c}}$ The meaning of the accuracy 
parameters is described
in Ref.\ [20] and are given in the order described in the user 
manual of Ref.\ [19].\ All values are obtained using
29 {\bf{k}} points for the IBZ integration.

\pagebreak 

\caption{\label{tab2} Scalar-relativistic effects in the FCC crystals AgF,
AgCl, AgBr and Ag$^{\rm{a}}$}
\begin{tabular}{llllll}
\multicolumn{1}{l}{Property}   
&\multicolumn{1}{c}{Method}   
&\multicolumn{1}{l}{\hspace{1.5ex}AgF}     
&\multicolumn{1}{l}{\hspace{1.5ex}AgCl}    
&\multicolumn{1}{l}{\hspace{1.5ex}AgBr}    
&\multicolumn{1}{l}{\hspace{1.5ex}Ag$^{\rm{b}}$}    \\ 
\tableline
${a_{0}}^{\rm{c,d}}$ (pm)
           & rel.\ HF &\hspace{1.5ex}517.5   &\hspace{1.5ex}593.7   
	   &\hspace{1.5ex}620.7   &\hspace{1.5ex}445.6   \\
	   & nonrel.\ HF &\hspace{1.5ex}519.1   &\hspace{1.5ex}599.0   
	   &\hspace{1.5ex}626.0   &\hspace{1.5ex}454.6   \\ [2.0ex]
$V_{0}$ ($\cmmo$)
           & rel.\ HF &\hspace{1.5ex}20.8664 &\hspace{1.5ex}31.5077 
	   &\hspace{1.5ex}36.0048 &\hspace{1.5ex}13.3214 \\
	   & nonrel.\ HF &\hspace{1.5ex}21.0605 &\hspace{1.5ex}32.3591 
	   &\hspace{1.5ex}36.9350 &\hspace{1.5ex}14.1450 \\[2.0ex]
$E_{0}$ ($\kjmol$)
           & rel.\ HF&$-371.2$&$-385.9$&$-333.4$&$-53.0$ \\
	   & nonrel.\ HF&$-407.5$&$-417.7$&$-362.8$&$-39.4$ \\[2.0ex]
${B_{0}}^{\rm{d}}$ (kbar)
           & rel.\ HF&\hspace{1.5ex}459 (459)&\hspace{1.5ex}306 (309)
	   &\hspace{1.5ex}273 (272)&\hspace{1.5ex}298 (295)\\
	   & nonrel.\ HF&\hspace{1.5ex}460 (461)&\hspace{1.5ex}297 (299)
	   &\hspace{1.5ex}259 (259)&\hspace{1.5ex}322 (323)\\[2.0ex]
${B^{'}_{0}}^{\rm{d}}$ 
           & rel.\ HF& \hspace{1.5ex}4.09 (4.12) 
	   &\hspace{1.5ex}5.03 (5.06) &\hspace{1.5ex}5.69 (5.45)
	   &\hspace{1.5ex}1.40 (1.06) \\
	   & nonrel.\ HF&\hspace{1.5ex}4.26 (4.31)
	   &\hspace{1.5ex}4.65 (4.75) &\hspace{1.5ex}5.63 (5.52)
	   &\hspace{1.5ex}1.40 (1.18)  \\
\end{tabular}
 
\pagebreak

\noindent $^{\rm{a}}$ Lattice parameter $a_{0}$, molar Volume $V_{0}$,
molar Hartree--Fock (HF) binding energy $E_{0}$ with reference
to the HF ground-state energies of the free atoms and to the
primitive unit cell, bulk modulus $B_{0}$ and pressure
derivative $B_{0}^{'}$.\ The relativistic all-electron CRYSTAL88
HF calculation using  $H^{{\rm{sf}}_{1}}_{+}$ (rel.\ HF)
is compared with a corresponding non-relativistic all-electron
CRYSTAL88 HF calculation (nonrel.\ HF).

\noindent $^{\rm{b}}$ Comparing non-relativistic and scalar-relativistic
results of Ref.\ [17] scalar-relativistic effects lead to a bond length
contraction (nearest-neighbor distance) of 5.7 pm (1.9\%), to a bond
stabilization of $-31.8\hspace{1.0ex}\kjmol$ (11.0\%)
and to an increase of the bulk modulus of
210 kbar (19.8\%) 
for Ag.\
The molar formation energy in Ref.\ [17] is calculated with reference to
the ground-state energy of the free silver atom and to the 
Wigner--Seitz cell volume which is identical to those of the 
primitive unit cell.\
The corresponding values for Ag in Table III are:
bond length contraction 6.4 pm (2.0\%), bond stabilization
$-13.6\hspace{1.0ex}\kjmol$ (34.5\%),
decrease in the bulk modulus 24 kbar (7.5\%)
(percentage with reference to non-relativistic
values).\

\noindent $^{\rm{c}}$ Parameters of the restart option GEOM of 
the CRYSTAL88 program are for AgF (5.00), AgCl (5.90),
AgBr (6.20) and Ag (4.45).\ The accuracy parameters are
the same as in Table II.

\noindent $^{\rm{d}}$ Ranges of bond distances used for fitting the EOS:
AgF [490 -- 540], AgCl [560 -- 630],
AgBr [605 -- 670] and Ag [425 -- 510] (all values in pm).\
The ranges have been chosen large enough to ensure numerical
stability, and small enough to exclude inappropriately
described regions of the potential curve at large bond distances,
taking the inadequacy of the closed-shell HF approximation
for near degenerate situations into account.\
Different EOSs [27] lead to
same results for $a_{0}$ -- otherwise values with an EOS of 
Birch [27]
are given in parentheses.\ 
All mean square deviations of the fits
are less than $0.6 \times 10^{-7}$.

\pagebreak 

\caption{\label{tab3} Crystal properties of FCC AgCl at various levels of
theory$^{\rm{a}}$}
\begin{tabular}{lcccccc}
Method & $E^{\rm{tot}}$ (a.u.) & $E_{0}$ ($\kjmol$) & $a_{0}$ (pm) &
$V_{0}$ ($\cmmo$) & $B_{0}$ (kbar) & $B^{'}_{0}$ \\
\tableline
nonrel.\ HF$^{\rm{b,f}}$ & $-5656.6014$ & $-417.2$ & 599.0 
& 32.3591 & 297 (299) & 4.65
(4.75)  \\ 
rel.\ HF$^{\rm{c,f}}$ & $-5771.3441$ & $-385.7$ & 593.7 
& 31.5077 & 306 (309) & 5.03 
(5.06) \\ 
rel.\ ECP$^{\rm{d,f}}$ & $\hspace{1.0ex}-159.7086$ & $-376.5$ & 594.3 
&  31.6000 & 307 (312)&
5.38 (5.44) \\
rel.\ ECP$^{\rm{e}}$& $\hspace{1.0ex}-159.708\hspace{1.0ex}$
& $-375.4$ &
594.0 & 31.5532 & 330 \hspace{5.0ex} & \\
DFT1$^{\rm{g}}$ & $\hspace{1.0ex}-160.656\hspace{1.0ex}$ & $-554.0$ &
570.0 & 27.8810 & 510 \hspace{5.0ex} & \\
DFT2$^{g}$ & $\hspace{1.0ex}-160.870\hspace{1.0ex}$ & $-514.6$ & 
561.0 & 26.5811& 560 \hspace{5.0ex} & \\
exp.$^{\rm{h}}$ &          & $-519.8$ & 551.0 
& 25.1848$^{\rm{i}}$ & 535 [417] &
[7.00] \hspace{5.0ex}
\end{tabular}

\pagebreak 

\pagestyle{empty}

\noindent $^{\rm{a}}$ Total energy $E^{\rm{tot}}$ with reference to
the primitive unit cell, molar formation
energy $E_{0}$ with reference to the ground-state energies
of the free atoms and to the primitive unit cell,
lattice parameter $a_{0}$, molar volume $V_{0}$,
isothermal bulk modulus $B_{0}$ and 
pressure derivative $B^{'}_{0}$.

\noindent $^{\rm{b}}$ Non-relativistic all-electron Hartree--Fock
(HF) calculation using the CRYSTAL88 program.

\noindent $^{\rm{c}}$ Relativistic all-electron Hartree--Fock (HF)
calculation
using the CRYSTAL88 program with $H^{{\rm{sf}}_{1}}_{+}$.

\noindent $^{\rm{d}}$ Relativistic effective-core potential (rel.\ ECP)
calculation using the 
CRYSTAL92 program and ECPs
for Ag and Cl given in Ref.\ [33].

\noindent $^{\rm{e}}$ Relativistic effective-core potentials (rel.\ ECP)
results calculated in Ref.\ [29]
using nearly the same ECPs
as in footnote d.\
(For Ag a (5s5p4d)/[3s3p1d] GTF basis set is used
instead of a (5s5p4d)/[3s3p2d] GTF
basis set with
$\xi=0.2108$ for 4d instead of $\xi=0.2079$ for 4d).

\noindent $^{\rm{f}}$ See footnote d of Table III.

\noindent $^{\rm{g}}$ Correlation contribution calculated
in Ref.\ [29] with a posteriori
density functionals of Colle and Salvetti (DFT1)
and Perdew (DFT2).

\noindent $^{\rm{h}}$ Estimated experimental values (see
Ref.\ [29] and references therein).\
The estimated values for $E_{0}$, $a_{0}$ and $B_{0}$
are an extrapolation of experimental results to T=0 K
including a zero-point energy correction.\
All values are zero-pressure
values.\ A zero-pressure value of the pressure
derivative $B^{'}_{0}$ for the B1 phase of AgCl at T=0 K
is not available.\ The estimated zero-pressure value
$B_{0}$ at room temperature and the corresponding 
pressure derivative $B^{'}_{0}$ (see Ref. [34])
are given in parentheses.

\noindent $^{\rm{i}}$ $V_{0}$ is deduced from $a_{0}$.

\pagestyle{empty}

\end{table}
\pagestyle{empty}
}

\pagestyle{empty}
\pagebreak

\pagestyle{empty}
\begin{figure}[ht]
\vspace*{6.0ex}
\epsfxsize = 15 cm
\leavevmode\epsffile{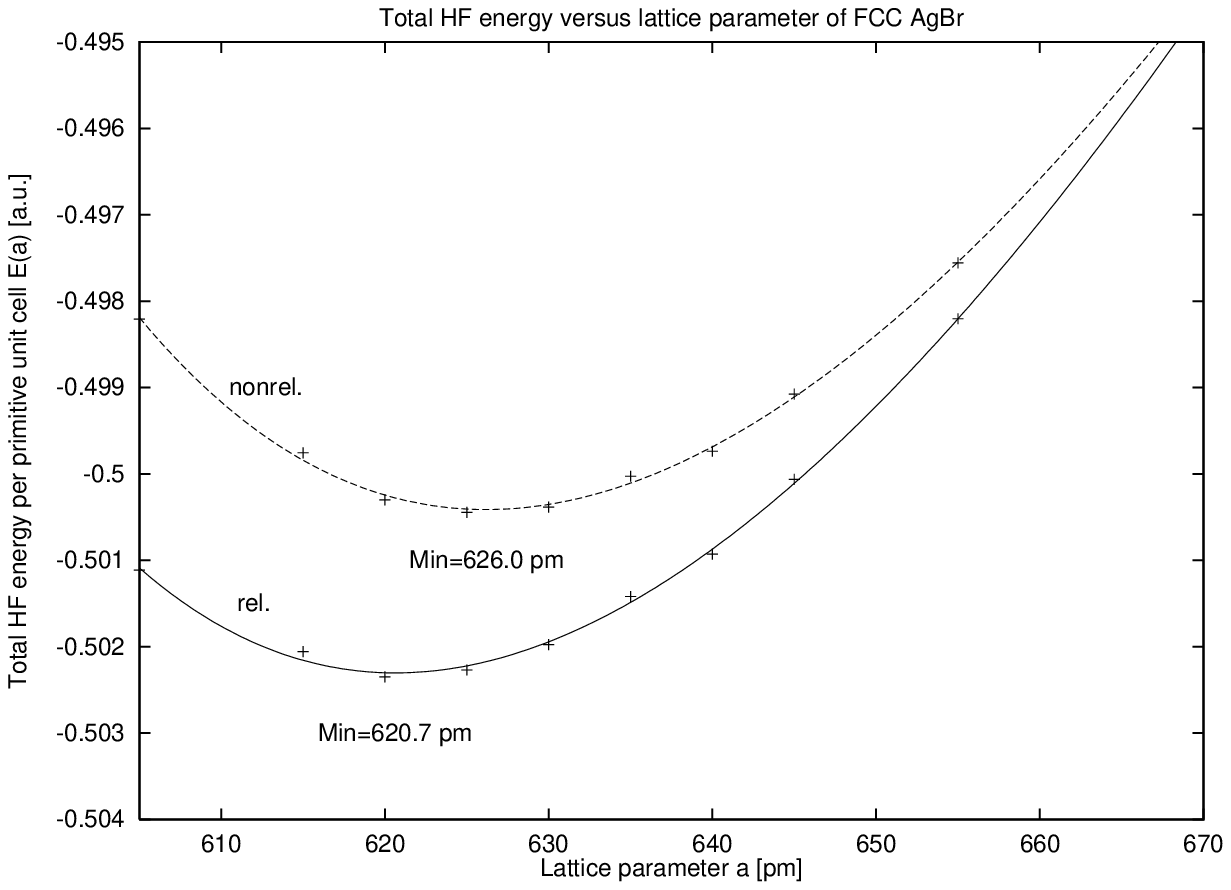}
\vspace*{6.0ex}
\caption{Total Hartree--Fock (HF) energy
versus lattice parameter of FCC AgBr.\
The fits using an EOS given by Murnaghan [27]
are shifted by 7769.0 au (nonrel.) and by
7914.165  au (rel.).\ Abbreviations used: rel.\
(relativistic CRYSTAL88 HF calculation using $H^{{\rm{sf}}_{1}}
_{+}$), nonrel.\ (non-relativistic CRYSTAL88 HF calculation).}
\label{abv}

\pagebreak

\pagestyle{empty}
\vspace*{6.0ex}
\epsfxsize = 15 cm
\leavevmode\epsffile{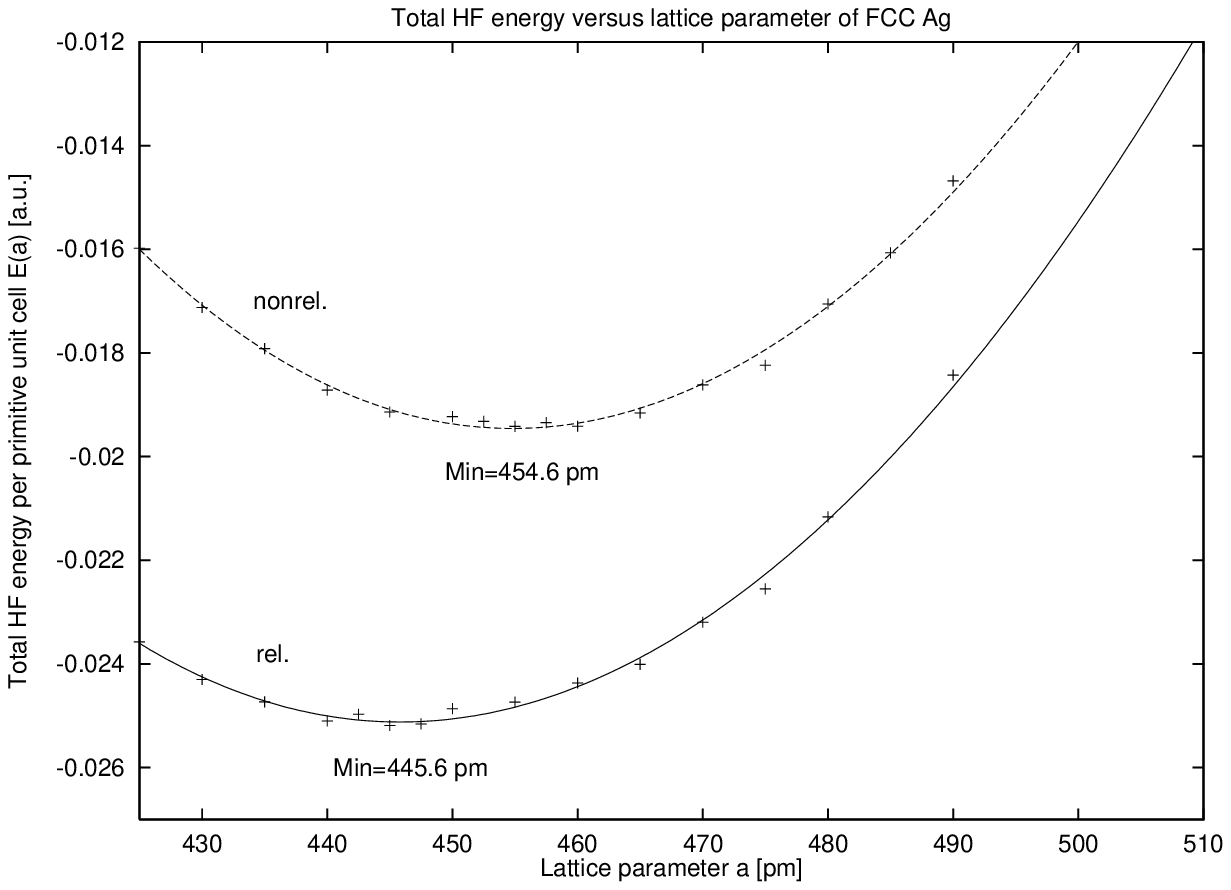}
\vspace*{6.0ex}
\caption{Total Hartree--Fock (HF) energy
versus lattice parameter of FCC Ag.\
The fits using an EOS given by Murnaghan [27]
are shifted by 5197.0 au (nonrel.) and by
5310.35  au (rel.).\ Abbreviations used: rel.\ 
(relativistic CRYSTAL88 HF calculation using $H^{{\rm{sf}}_{1}}_{+})$,
nonrel.\ (non-relativistic CRYSTAL88 HF calculation).}
\label{agv}

\pagebreak

\pagestyle{empty}
\vspace*{6.0ex}
\epsfxsize = 15 cm
\leavevmode\epsffile{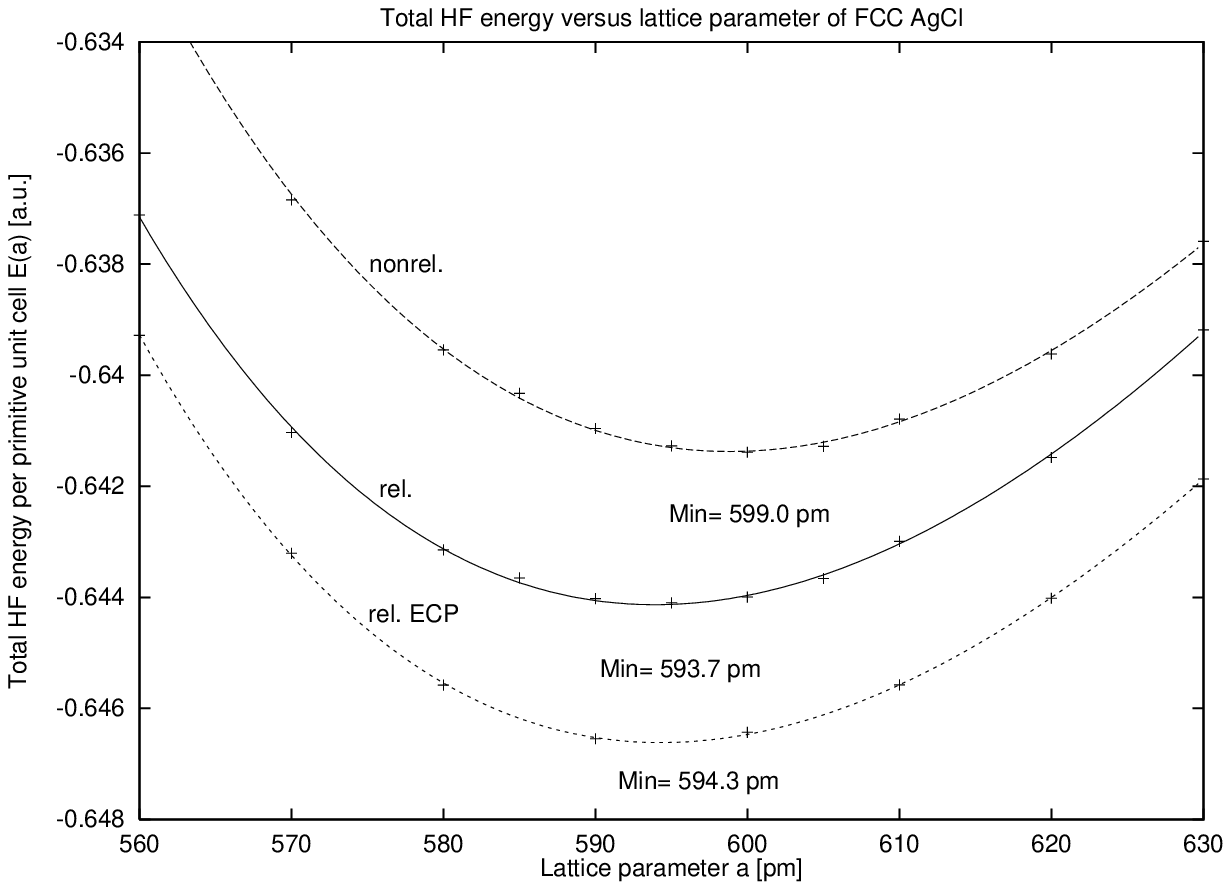}
\vspace*{6.0ex}
\caption{Total Hartree--Fock (HF) energy 
versus lattice parameter of FCC AgCl.\
The fits using an EOS given by Murnaghan [27]
are shifted by 5656.0 au (nonrel.),
by 5770.7 au (rel.) and by 159.062 au (rel.\ ECP.).\
Abbreviations used: rel.\ (relativistic CRYSTAL88 HF calculation
using $H^{{\rm{sf}}_{1}}_{+}$),
nonrel.\ (non-relativistic CRYSTAL88 HF calculation),
rel.\ ECP (relativistic effective-core potential
CRYSTAL92 HF calculation [33]).
}
\label{alv}
\end{figure}
\end{document}